\documentclass[final,5p,times,twocolumn]{elsarticle}

\usepackage{graphicx}
\usepackage{amssymb}
\usepackage{amsmath}

\usepackage[usenames]{color}
\definecolor{darkblue}{rgb}{0,0.08,0.45}
\usepackage[colorlinks=true,linkcolor=darkblue,anchorcolor=darkblue,citecolor=darkblue,urlcolor=darkblue,pdfstartview=FitH]{hyperref}
\usepackage{algorithm}
\usepackage{listings}

\definecolor{dkgreen}{rgb}{0,0.6,0}
\definecolor{gray}{rgb}{0.5,0.5,0.5}
\definecolor{mauve}{rgb}{0.58,0,0.82}

\lstdefinelanguage{MyPython}[]{python}
{
  basicstyle=\small,
  showspaces=false,
  showtabs=false,
  rulecolor=\color{black},
  tabsize=2,
  captionpos=b,
  breaklines=true,
  breakatwhitespace=false,
  commentstyle=\color{dkgreen},
  stringstyle=\color{mauve},
  numberstyle=\bfseries
  escapeinside={\%*}{*)},
  deletekeywords=[2]{set,all}
}

\floatname{algorithm}{Listing}

\journal{arxiv}

\begin{document}

\begin{frontmatter}

\title{HOOMD-blue: A Python package for high-performance molecular dynamics and hard particle Monte Carlo simulations}

\author[chem]{Joshua A. Anderson}
\author[chem]{Jens Glaser}
\author[chem,mse,bio]{\texorpdfstring{Sharon C. Glotzer\corref{cor1}}{Sharon C. Glotzer}}
\ead{sglotzer@umich.edu}
\cortext[cor1]{Corresponding author}

\address[chem]{Department of Chemical Engineering, University of Michigan, Ann Arbor,
Michigan, 48109, USA}
\address[mse]{Department of Material Science and Engineering, University of Michigan, Ann Arbor, Michigan, 48109, USA}
\address[bio]{Biointerfaces Institute, University of Michigan, Ann Arbor, Michigan, 48109, USA}

\begin{abstract}
HOOMD-blue is a particle simulation engine designed for nano- and colloidal-scale molecular dynamics and hard particle Monte Carlo simulations.
It has been actively developed since March 2007 and available open source since August 2008.
HOOMD-blue is a Python package with a high performance C++/CUDA backend that we built from the ground up for GPU acceleration.
The Python interface allows users to combine HOOMD-blue with with other packages in the Python ecosystem to create simulation and analysis workflows.
We employ software engineering practices to develop, test, maintain, and expand the code.
\end{abstract}

\begin{keyword}
Python \sep molecular dynamics \sep Monte Carlo \sep Molecular simulation \sep GPU \sep CUDA
\end{keyword}

\end{frontmatter}

\section{Introduction}
\label{sec:introduction}

Molecular, nano-, and colloidal-scale simulations are powerful tools to probe the structure and dynamics of materials.
Simulations can offer insight to the fundamental physics of a phenomenon and be used to efficiently scan parameter space to find promising structures, properties or behavior.
Molecular dynamics (MD) is commonly used in the biomolecular community with long established codes such as AMBER~\cite{Amber2018}, GROMACS~\cite{GROMACS}, and NAMD~\cite{Phillips2005}.
LAMMPS~\cite{Plimpton1995} is a general purpose MD engine with many capabilities designed for materials science applications, and is also capable of biomolecular simulations.
Monte Carlo (MC) simulations of molecular systems are well suited for determining phase equilibria and made possible by codes like Cassandra~\cite{Shah2017}, GOMC~\cite{Nejahi2019}, and MCCCS Towhee~\cite{Martin2013}.
Recent codes like FEN ZI~\cite{Ganesan2011, Taufer2013}, HOOMD-blue~\cite{Anderson2008a}, and OpenMM~\cite{Eastman2010} were developed around new functionalities or use cases not possible with the established codes.

Over the past decade, Python has become a popular scripting language for scientific computing in general~\cite{Millman2011} and the molecular simulation community in particular.
Many Python based tools are now available for system initialization, trajectory analysis, and workflow management such as mBuild~\cite{Klein2016}, MDAnalysis~\cite{Michaud-Agrawal2011}, MDTraj~\cite{McGibbon2015}, pysimm~\cite{Fortunato2017}, and signac~\cite{Adorf2018d}.
Of the simulation engines mentioned, only HOOMD-blue and OpenMM provide first-class Python application programming interfaces (APIs).

This paper describes HOOMD-blue v2.6.
HOOMD-blue is a particle simulation engine designed for nano- and colloidal-scale simulations.
As a general-purpose tool HOOMD-blue is capable of standard MD and biomolecular simulations, but we focus our efforts on providing unique functionality that is not already available in other codes.

HOOMD-blue has been under development for more than 10 years (\autoref{sec:history}) as an open source code.
It provides MD and hard particle MC capabilities (\autoref{sec:capabilities}).
We implement HOOMD-blue as a Python package (\autoref{sec:python}) that seamlessly interoperates with the
scientific Python ecosystem.
The high level Python interface abstracts a high performance backend (\autoref{sec:performance}) that executes simulations on one or many GPUs or CPUs.
As an open source project (\autoref{sec:extendable}), users can extend the code with  new models and methods to enable their research.
We employ software engineering (\autoref{sec:software_engineering}) practices to design, implement, and test the code.

\section{History}
\label{sec:history}

HOOMD began development in March 2007 at Iowa State University.
The initial implementation consisted of algorithms and data structures demonstrating high performance MD on the GPU for coarse-grained polymer simulations~\cite{Anderson2008a}, implemented in C++ and CUDA.
The second open source release in 2008 added a Python interface.
Research groups discovered HOOMD, contributed new functionalities to the code, and published research papers using it~\cite{Phillips2011,Levine2011,Morozov2011,Nguyen2011,LeBard2012}.
In August 2009, HOOMD development moved to the University of Michigan and the software became HOOMD-blue (HOOMD, blue edition).
Development on the project has continued with more than 10,000 commits by 68 contributors since inception.
Two major milestones in HOOMD-blue development were the releases of v1.0 in 2014, which added MPI domain decomposition~\cite{Glaser2014c}, and v2.0 in 2016 which added Monte Carlo~\cite{Anderson2016a,Glaser2015c} and discrete element method MD~\cite{Spellings} for hard shapes.
In addition to these major developments, the code has grown organically with new capabilities and performance improvements through a process of lazy refactoring~\cite{Adorf2018a}.
As of September 2019, we are aware of 298 peer-reviewed research articles that made use of HOOMD-blue.

\section{Capabilities}
\label{sec:capabilities}

\subsection{Molecular dynamics}

\begin{algorithm}[t]
\caption{Molecular dynamics simulation}
\label{alg:md_script}

\begin{lstlisting}[language=MyPython]
from hoomd import *
from hoomd import md
# place particles
context.initialize()
unitcell=lattice.sc(a=2.0, type_name='A')
init.create_lattice(unitcell, n=10)
# define Lennard-Jones interactions
nl = md.nlist.cell()
lj = md.pair.lj(r_cut=2.5, nlist=nl)
lj.pair_coeff.set('A', 'A',
                  epsilon=1.0,sigma=1.0)
# NVT integration
all = group.all();
md.integrate.mode_standard(dt=0.005)
nvt = md.integrate.nvt(group=all, kT=1.2,
                       tau=1.0)
nvt.randomize_velocities(seed=1)
# run the simulation
run(10e3)
\end{lstlisting}
\end{algorithm}

HOOMD-blue has MD integrators for many different thermodynamic ensembles including NVE, NVT, NPH, NPT, Langevin dynamics, Brownian dynamics, and dissipative particle dynamics, and also supports FIRE~\cite{Bitzek2006} energy minimization.
The NVT, NPH, and NPT integrators are based on the Martyna-Tobias-Klein method~\cite{Martyna1994a}.
All of these support the integration of rotational degrees of freedom directly, and are able to couple a single thermostat to a system consisting of particles with and without rotational degrees of freedom.
HOOMD-blue also implements the multi-particle collision dynamics solvent model~\cite{Howard2018a}.
Users can apply different integrators to distinct subsets of the system.

The general-purpose MD engine in HOOMD-blue can apply many different types of forces to particles.
Users can employ any number of these to achieve the desired model.
Anisotropic potentials treat particles with extended shape and produce both forces and torques on particles.
HOOMD-blue implements DEM potentials for faceted shapes~\cite{Spellings}, Gay-berne ellipsoids, and dipole potentials.
Composite particle constraints connect many constituent particles so that they move as a rigid body~\cite{Nguyen2011, Glaser2019a}.
HOOMD-blue supports the active matter community with an active force module that can apply constant magnitude forces or torques to particles.
HOOMD-blue also supports commonly used pair, bond, angle, dihedral, improper, special pair potentials, and PPPM electrostatics~\cite{LeBard2012}.
We provide a wide variety of pair potentials utilized in different fields including Buckingham, DLVO, DPD, Lennard-Jones, Gaussian, Mie, WCA, Yukawa, and others.
Users can also develop and test custom potentials with tabulated pair, bond, angle and dihedral potentials.
HOOMD-blue includes EAM~\cite{Morozov2011, Yang2018}, Tersoff, and square density many-body potentials as well as periodic, electric field, constant force, and wall external potentials.
Users can fix the bond length between pairs of particles with distance constraints~\cite{Yoneya1994,Yoneya2001}, or constrain particles to the surface of a sphere or an ellipsoid.

\subsection{Monte Carlo}

\begin{algorithm}[t]
\caption{Hard particle Monte Carlo simulation}
\label{alg:mc_script}

\begin{lstlisting}[language=MyPython]
from hoomd import *
from hoomd import hpmc
# place particles
context.initialize()
unitcell=lattice.sc(a=2.0, type_name='A')
init.create_lattice(unitcell, n=10)
# hard particle Monte Carlo
mc = hpmc.integrate.convex_polyhedron(
                    d=0.1, a=0.1, seed=2)
cube_verts = \
  [[-0.5,-0.5,-0.5], [0.5,-0.5,-0.5],
   [-0.5,-0.5, 0.5], [0.5,-0.5, 0.5],
   [-0.5, 0.5,-0.5], [0.5, 0.5,-0.5],
   [-0.5, 0.5, 0.5], [0.5, 0.5, 0.5]]
mc.shape_param.set('A',
                    vertices=cube_verts)
# run the simulation
run(10e3)
\end{lstlisting}
\end{algorithm}

In addition to MD simulations, HOOMD-blue can perform hard particle MC simulations~\cite{Anderson2016a} with the HPMC component.
We have implemented a wide variety of shape classes, including spheres, disks, unions of spheres, convex spheropolygons, simple polygons, ellipsoids, convex spheropolyhedra, unions of convex spheropolyhedra, faceted spheres, and general triangle meshes.
All particles in a simulation must be of the same shape class, but may be of different particle types, where each particle type has separate shape parameters.
User-defined wall constraints confine particles to particular regions of space.
HPMC implements trial moves that enable NVT, NPT, grand, and Gibbs ensembles.
An implicit depletant algorithm~\cite{Glaser2015c} enables efficient simulations of colloids with depletants.
HPMC can sample the simulation pressure in NVT ensembles~\cite{Eppenga1984,Brumby2011,Anderson2016a} and the free volume available to the system.
We have also implemented the Frenkel-Ladd free energy method~\cite{Frenkel1984b}.

In hard particle Monte Carlo, the energy of the system is infinite when two particles overlap or zero when there are no overlaps.
There are research applications that apply attractive patchy interactions in addition to the hard particle core.
Each project customizes the form of this enthalpic interaction to the specific research question.
HOOMD provides flexibility to the user while maintaining high performance using just-in-time compilation.
The user provides a C++ code snippet in their script, which HPMC compiles at runtime with Clang~\cite{Clang} and executes it with LLVM~\cite{Lattner2004} when needed to determine the energy of each interaction.
We provide hooks for cutoff pair potentials, cutoff pair potentials evaluated at the points of a sphere union shape, and external potentials applied to each particle.

\section{Python package}
\label{sec:python}

HOOMD-blue is a Python package.
Through the imperative Python API, a user can configure which capabilities are enabled, set parameters, and control the progression of the simulation run.
\autoref{alg:md_script} performs a simple MD simulation.
It uses \texttt{init.create\_lattice} to initialize a simple cubic lattice of particles, \texttt{md.pair.lj} to define the Lennard-Jones particle pair interaction potential, and \texttt{md.integrate.nvt} to employ NVT ensemble integration.
\autoref{alg:mc_script} performs a simple MC simulation of hard cubes.
It uses \texttt{hpmc.integrate.convex\_polyhedron} to specify HPMC integration of the convex polyhedron shape class and sets the parameters for type \emph{A} to the vertices of a cube.
In both examples \texttt{run} executes the simulation for the given number of steps.

Users can use the Python API in a script, via job queue submission, inside Jupyter notebooks, or by any other method one can use a Python package.
For example, HOOMD-blue can easily be used for workflow steps in the \emph{signac} framework~\cite{Adorf2018d}.
Options passed to the device context control whether the simulation executes on the CPU or the GPU.
Without modifications to the script, a user may execute a script on many CPUs or GPUs by launching it as an MPI parallel job or on GPU nodes with NVLINK (such as OLCF Summit) with the multiple GPU device option \texttt{--gpu=0,1,2}.

We provide complete documentation~\cite{HOOMD-documentation2019} for HOOMD-blue, including installation instructions, tutorials in Jupyter notebooks, overview documentation of general concepts, and a complete listing of every API call along with the options and parameters they accept.
Users can browse static copies of the tutorials online or download the notebooks and execute them with Jupyter.
Jupyter notebooks provide a mixture of formatted text description, code, and output.
The HOOMD-blue example notebooks examine the simulation results and show relevant output, including plots of system quantities vs time and trajectory visualizations.
When running the notebooks, users can modify parameters or introduce new commands, then re-execute the notebook and see how the output changes.
HOOMD-blue's user and developer community is available to answer questions on the hoomd-users mailing list, which has 521 members as of September 2019.

HOOMD-blue runs on Linux and macOS.
We provide binary packages on the \emph{conda-forge}~\cite{condaforge} Anaconda channel and also in Docker and Singularity~\cite{Kurtzer2017} images.
The Anaconda packages provide an installation mechanism suitable for testing jobs on a laptop or workstation.
However, there are limitations that prevent Anaconda packages from taking full advantage of resources on HPC clusters.
We provide Singularity images with performance optimized builds of HOOMD for a number of national HPC resources, including PSC Bridges, SDSC Comet, and TACC Stampede2.
Users can also build HOOMD-blue from source using CMake, numpy, and Python.
NVIDIA CUDA is optional, but required to enable GPU acceleration.
An MPI library is optional, but required to enable execution in parallel on multiple nodes.

\section{Performance}
\label{sec:performance}

We optimize all capabilities in HOOMD-blue extensively so that it runs nano- and colloidal-scale simulations as fast as possible.
With very few exceptions, all of the time-consuming operations in HOOMD-blue can execute on the GPU, including all of the data structures, force evaluations, integrators, and MPI communication.
We have also parallelized the majority of HOOMD-blue capabilities with MPI for execution on multiple GPUs or multiple CPU cores.
The exceptions are cases where particular methods are typically used for simulations where GPUs and/or MPI simulations are not necessary or code paths that are rarely called.
The components that lack GPU support in v2.6 are external fields and user-defined pair potentials in HPMC, and the temperature rescale and zero momentum updaters in MD.
The components that lack MPI support in v2.6 are active forces, the IMD (interactive MD) communication protocol, the Berendsen integrator, and ellipsoid constraints.

Simulation performance is highly dependent on the type of simulation, particle sizes, force fields, system density, and other parameters.
Typical research-relevant MD simulations execute an order of magnitude faster on a single GPU than on all cores of a single CPU socket for system sizes of four thousand particles per GPU or more~\cite{Glaser2014c}.
Over the years, we have continually re-tuned and rewritten the MD CUDA code for each new generation of GPU hardware.
The current version of the code (v2.6) utilizes many optimization techniques standard in the CUDA developer community, including multiple threads per particle with warp-level reductions, atomic operations to build cell lists, auto-tuning kernel parameters, and others which readers can find in the HOOMD-blue source code.
Recently, we added support for improved intra-node scaling to many GPUs using NVIDIA's NVLINK technology~\cite{Glaser2019}.

HOOMD-blue includes unique performance optimizations.
In colloidal systems with large and small particle sizes, cell-list based neighbor list algorithms do not operate efficiently.
The cell list must either be sized to the largest particle, or many hundreds of cells must be traversed to find tens of neighbors.
Bounding volume hierarchy (BVH) data structures, commonly used in computer graphics and video games, dynamically adapt to the density fluctuations with minimal memory usage and high performance.
HOOMD-blue provides a BVH method to build neighbor lists with \texttt{nlist.tree}, contributed by Michael Howard~\cite{Howard2016}, which is faster than the cell list for size ratios of 2:1 or greater.
Howard is currently working to optimize this code path further with the goal to make it faster than the standard cell list in all cases~\cite{Howard2019}.
Users choose which neighbor list implementation to use in their simulations.

Hard particle Monte Carlo (HPMC) simulations also execute an order of magnitude faster on a single GPU than on all cores of a single CPU socket, but only when system sizes are larger than tens of thousands of particles per GPU.
We are actively working on further optimizing this GPU code path for hard particle simulations smaller than this.
We have also spent considerable effort optimizing HPMC's CPU code path.
HPMC uses BVH data structures and CPU vector intrinsics, and executes many trial moves in parallel to attain the best possible performance on the CPU.
See ref.~\cite{Anderson2016a} for complete details about the HPMC component of HOOMD-blue.

To put these general performance statements into context, we include benchmarks for two representative cases:
the Lennard-Jones liquid with $N=64,000$ particles at a number density of $0.382\sigma^{-3}$ in the NVT ensemble with $k_\mathrm{B}T/\varepsilon = 1.2$, $\delta t = 0.005\sqrt{m\sigma^2/\varepsilon}$ and a cutoff distance $r_\mathrm{cut}=3.0\sigma$~\cite{Anderson2008a}; and the hard hexagon hexatic phase with $N=1,048,576$ particles at a packing fraction of $\phi_\mathrm{P}=0.7$ in the NVT ensemble~\cite{Anderson2017}.

On a single 24-core Intel Xeon Platinum 8160 CPU (TACC Stampede2), HOOMD-blue v2.6.0 performs the MD Lennard-Jones liquid benchmark at $16.1\cdot10^6$ particle time steps per second and the HPMC hexagon benchmark at $20.8\cdot10^6$ trial moves per second.
On a single V100 GPU (PSC Bridges), performance increases to $275\cdot10^6$ million particle time steps per second for the Lennard-Jones liquid benchmark and $132\cdot10^6$ trial moves per second for the hexagon benchmark.

We recommend that users test their models with different numbers of CPU cores and GPUs so that they can make informed choices on the performance and efficiency tradeoffs specific to their systems.
We urge authors who wish to publish comparative benchmarks to build the latest version of HOOMD-blue and perform direct comparisons on the most current hardware available.
The performance numbers we include here are representative only of a single point in time.
Manufactures regularly produce new processors and developers frequently write new code performance optimizations.
As a case in point, compare the above \emph{full double precision} V100 performance of HOOMD-blue v2.6.0 to our 2008 publication, where HOOMD v0.6.0 performed the Lennard-Jones liquid benchmark \emph{in single precision} on a single G80 GPU at $12.9\cdot10^6$ particle time steps per second~\cite{Anderson2008a} and our 2015 publication where HOOMD v1.0.0 performed the Lennard-Jones benchmark (with $N=32,000$) \emph{in single precision} on a single K20 GPU (OLCF Titan) at $64\cdot10^6$ particle time steps per second~\cite{Glaser2014c}.

\section{Open source}
\label{sec:extendable}

HOOMD-blue is available open source under the permissive 3-clause BSD license.
To implement changes, users can fork the HOOMD-blue code and add new functionalities directly, or create a plugin in a separate code repository and link it to HOOMD-blue at build time.
Many users have published papers and software frameworks using HOOMD-blue with extensions they have developed.
Six recent examples of this include: raaSAFT, a framework enabling coarse-grained simulations based on the SAFT-$\gamma$ Mie force field~\cite{Ervik2017}; reverse non-equilibrium molecular dynamics simulation applied to transitions between lamellar orientations in shear flow~\cite{Schneider2018a}; protracted colored noise dynamics applied to linear polymer systems~\cite{Peters2018a}; epoxpy, a package that dynamically adds bonds during DPD simulations to model expoxy curing~\cite{Thomas2018a}; accurate hydrodynamic interactions to study how surface heterogeneity affects percolation and gelation of colloids~\cite{Wang2019b}; and Hoobas, a highly objected-oriented builder for molecular dynamics to generate complex initial conditions for polymer and DNA-coated nanoparticle systems~\cite{Girard2019a}.

As of September 2019, GitHub (\url{https://github.com/glotzerlab/hoomd-blue}) facilitates the open source development of HOOMD-blue (previously, we have hosted development on Assembla, Redmine, and Bitbucket).
The code repository houses the entire history of HOOMD-blue's development and allows many developers to simultaneously work on changes to the code.
Issue lists allow users to submit bug reports and track their progress and developers to track planned feature development.
We encourage contributions to HOOMD-blue from the community.
Pull requests allow the community to review and discuss proposed changes to the code.
We merge pull requests into the main line of development after they are reviewed and approved and pass all tests.

\section{Software engineering}
\label{sec:software_engineering}

HOOMD-blue v2.6 consists of 173,662 lines of code.
While 68 individuals have contributed to the project over its lifetime, at any one time a group of 2-3 developers maintains the code and improves core functionalities.
HOOMD-blue developers are researchers and do not develop full-time.
We automate processes and carefully consider library dependencies, tools, and design patterns in order to provide the highest quality code that will remain stable over a long time period with a minimum of maintenance effort.
For example, HOOMD-blue's simulation engine is written with a modular, isolated, object oriented design.
Each capability of the code is implemented in a separate class that communicates with the core particle data structures and only with other classes as necessary.

We implement simulation algorithms in C++ and CUDA, with MPI for domain decomposition.
CUDA and MPI are optional, but are needed to provide GPU and parallel runs, respectively.
The user-facing side HOOMD-blue is written in Python.
We use the pybind11~\cite{pybind11} library to interface Python and C++ classes, and the CMake system to configure builds.
We also utilize a number of header-only libraries which we embed with submodules so that users and developers do not need to separately download them.
HOOMD-blue's documentation contains the full list of all libraries utilized and their corresponding license notices.

\subsection{Testing}

We employ white box unit testing for every module in HOOMD-blue.
Each test is designed with knowledge of how the module is implemented and sufficient cases are tested so as to exercise all of the possible code paths that the module can take.
The automated unit tests can easily be executed for any build of HOOMD-blue to validate that all capabilities are operating correctly.

For example, a unit test for the Lennard-Jones pair force class creates a system with six particles in it.
Some of these interact across periodic boundary conditions while others interact directly.
It then sets the potential parameters $\varepsilon$, $\sigma$, and $r_\mathrm{cut}$ to different values and verifies that the correct force, energy, and virial are computed each time.
We compute reference values independently (e.g. with a calculator) for individual unit tests.
A test passes only when all computed values are within a tolerance of the reference.

We also perform black box system integration tests to ensure that modules work together correctly.
These validation tests assume no implementation specific knowledge.
The tests enter simulation parameters in a job script and analyze output files the same way a user would run HOOMD-blue.

Automated validation tests must meet the following requirements.
The test simulation must be long enough to sample an average value with reasonable error bars, but short enough that they complete in a reasonable amount of time (ideally, individual tests should complete in less than 10 minutes).
When available, we validate equations of state against published reference values.
For example, we use the pressure and density of the hard disk fluid to validate HPMC~\cite{Bernard2011} and the NIST standard simulation reference~\cite{NISTSimulationReference2017} to validate Lennard-Jones simulations in both HPMC and MD.
In other cases, such references are not available in the literature and we instead cross-validate multiple code paths in HOOMD-blue.
For example, one validation test compares MD and MC simulations of WCA dimers, both performed by HOOMD-blue.

\subsection{Continuous integration}

We employ continuous integration practices to ensure code quality.
Scripts trigger on every commit to the HOOMD-blue source code repository and on pull requests.
These scripts compile that commit with a selection of compilers, CUDA versions, Python versions, CMake versions, LLVM versions, and other build options.
Then the script runs the unit tests.
Selected builds also run the longer validation tests.
All of the build output is captured and any failing builds or failing test are flagged.
The continuous integration testing system provides developers with feedback on the test outcomes.

As of September 2019, we build Docker containers to provide the build and test environment and execute tests on Microsoft Azure Pipelines~\cite{Azure2019}.
We execute CPU tests on Microsoft-Hosted cloud agents and GPU tests on self-hosted agents we run locally.
We also utilize the readthedocs~\cite{readthedocs} service to automatically build and host HOOMD-blue's documentation when new commits are pushed to the repository.

\section{Conclusions}
\label{sec:conclusions}

HOOMD-blue is a particle simulation engine designed for nano- and colloidal-scale simulations.
It was built from the ground up for GPU acceleration, and has been actively developed since March 2007 with more than 10,000 commits and 68 contributors.
As a general-purpose code, it is used by researchers the fields of colloidal self-assembly, active matter, coarse grained polymers, and others.

We develop the open source HOOMD-blue as a Python package with a high performance C++/CUDA backend.
HOOMD-blue's Python API allows users to define and execute simulations and combine them with other tools in the scientific Python ecosystem.
We optimize all core functionalities to obtain the fastest possible performance on whatever architecture (CPU or GPU) is best suited to the problem.
This requires us to continually refactor and rewrite the core kernels to ensure HOOMD-blue performs well on the latest hardware.
As an open source package, users can modify and extend the code to meet their specific needs.

We employ software engineering practices including unit tests, system validation tests and continuous integration to ensure that the code operates correctly.
Collaborative development and code review practices to ensure code meets standards set by the developers and to give the community an opportunity to provide input on the development process.
HOOMD-blue is open source and we welcome contributions of new capabilities of interest to a wide audience of users.

We are impressed by the amazing science the research community is able to accomplish using HOOMD-blue.
We hope that more of this great work continues, and that more developers contribute to make HOOMD-blue a better simulation code for the community.

\section{Acknowledgments}
\label{sec:acknowledge}

Initial HOOMD development (v0.6-v0.8) was supervised by Alex Travesset and funded by NSF through Grant DMR-0426597 and by DOE through the Ames lab under Contract No. DE-AC02-07CH11358.
HOOMD-blue development has been supported by the DOD/ASD(R\&E) under Award No. N00244-09-1-0062 (2009-2014, early design and implementation, v0.9 – v1.x) and the National Science Foundation, Division of Materials Research Award \# DMR 1409620 (2014-2018, especially DEM and HPMC capabilities in v2.x). Software was validated and benchmarked on the Extreme Science and Engineering Discovery Environment (XSEDE)~\cite{Towns2014}, which is supported by National Science Foundation grant number ACI-1053575 (XSEDE award DMR 140129); on resources of the Oak Ridge Leadership Computing Facility which is a DOE Office of Science User Facility supported under Contract No. DE- AC05-00OR22725; and through computational resources and services provided by Advanced Research Computing at the University of Michigan, Ann Arbor. Hardware provided by NVIDIA Corp. is gratefully acknowledged.  Any opinions, findings, and conclusions or recommendations expressed in this publication are those of the author(s) and do not necessarily reflect the views of the DOD/ASD(R\&E).

We would like to thank all HOOMD-blue contributors: Carl Simon Adorf, Khalid Ahmed, James Antonaglia, Steve Barr, Joseph Berleant, Isaac Bruss, Chengyu Dai, Kevin Daly, Avisek Das, Bradley Dice, Paul Dodd, Chrisy Du, Åsmund Ervik, Jenny Fothergill, Grey Garrett, Eric Harper, Mike Henry, Michael Howard, Alexander Hudson, M. Eric Irrgang, Eric Jankowski, Kwanghwi Je, Bjørnar Jensen, Christoph Junghans, Aaron Keys, Christoph Klein, Axel Kohlmeyer, Kevin Kohlstedt, David LeBard, Andrew Mark, Ryan Marson, Tim Moore, Shannon Moran, Igor Morozov, Pavani Medapuram Lakshmi Narasimha, Richmond Newman, Trung Dac Nguyen, Sam Nola, Antonio Osorio, Carolyn Phillips, James Proctor, Cong Qiao, Vyas Ramasubramani, Malcolm Ramsay, Sumedh R. Risbud, Luis Y. Rivera-Rivera, Ludwig Schneider, Benjamin Schultz, Peter Schwendeman, Wenbo Shen, Kevin Silmore, Rastko Sknepnek, Brandon Denis Smith, Ross Smith, Matthew Spellings, Ben Swerdlow, Erin Teich, Stephen Thomas, Alex Travesset, Alyssa Travitz, Greg van Anders, Bryan VanSaders, Lin Yang, Pengji Zhou, and William Zygmunt

\section{Data availability}

HOOMD-blue source code is available on github: https://github.com/glotzerlab/hoomd-blue

\bibliographystyle{elsarticle-num-names}
\bibliography{hoomd}

\end{document}